\documentclass[twocolumn]{aastex62}

\graphicspath{{Figures/}}

\received{}
\revised{}
\accepted{}
\submitjournal{AJ}

\shorttitle{RDI with Keck/NIRC2}
\shortauthors{Ruane et al.}

\begin{document}

\title{Reference star differential imaging of close-in companions and circumstellar disks with the NIRC2 vortex coronagraph at W.M. Keck Observatory}

\correspondingauthor{Garreth Ruane}
\email{gruane@caltech.edu}

\author[0000-0003-4769-1665]{Garreth Ruane}
\altaffiliation{NSF Astronomy and Astrophysics Postdoctoral Fellow}
\affil{Department of Astronomy, California Institute of Technology, 1200 E. California Blvd., Pasadena, CA 91125, USA}

\author[0000-0001-5172-4859]{Henry Ngo}
\affil{NRC Herzberg Astronomy and Astrophysics, 5071 West Saanich Road, Victoria, British Columbia, Canada}

\author[0000-0002-8895-4735]{Dimitri Mawet}
\affil{Department of Astronomy, California Institute of Technology, 1200 E. California Blvd., Pasadena, CA 91125, USA}
\affil{Jet Propulsion Laboratory, California Institute of Technology, 4800 Oak Grove Dr., Pasadena, CA 91109, USA}

\author[0000-0002-4006-6237]{Olivier Absil}
\altaffiliation{F.R.S.-FNRS Research Associate}
\affil{Space Sciences, Technologies, and Astrophysics Research (STAR) Institute, Universit\'{e} de Li\`{e}ge, Li\`{e}ge, Belgium}

\author[0000-0002-9173-0740]{\'{E}lodie Choquet}
\altaffiliation{Hubble Fellow}
\affil{Department of Astronomy, California Institute of Technology, 1200 E. California Blvd., Pasadena, CA 91125, USA}
\affil{Aix Marseille Univ., CNRS, CNES, LAM, Marseille, France}

\author{Therese Cook}
\affil{Department of Astronomy, California Institute of Technology, 1200 E. California Blvd., Pasadena, CA 91125, USA}
\affil{Department of Physics and Astronomy, University of California, Los Angeles, CA 90095, USA}

\author[0000-0003-2050-1710]{Carlos Gomez Gonzalez}
\affil{Universit\'{e}́ Grenoble Alpes, IPAG, F-38000 Grenoble, France}

\author{Elsa Huby}
\affil{LESIA, Observatoire de Paris, Universit\'{e} PSL, CNRS, Sorbonne Universit\'{e}, Univ. Paris Diderot, Sorbonne Paris Cit\'{e},\\5 place Jules Janssen, 92195 Meudon, France}

\author{Keith Matthews}
\affil{Department of Astronomy, California Institute of Technology, 1200 E. California Blvd., Pasadena, CA 91125, USA}

\author[0000-0001-6126-2467]{Tiffany Meshkat}
\affil{IPAC, California Institute of Technology, 1200 E. California Blvd., Pasadena, CA 91125, USA}

\author[0000-0003-2911-0898]{Maddalena Reggiani}
\affil{Space Sciences, Technologies, and Astrophysics Research (STAR) Institute, Universit\'{e} de Li\`{e}ge, Li\`{e}ge, Belgium}

\author{Eugene Serabyn}
\affil{Jet Propulsion Laboratory, California Institute of Technology, 4800 Oak Grove Dr., Pasadena, CA 91109, USA}

\author[0000-0003-0354-0187]{Nicole Wallack}
\affil{Division of Geological and Planetary Sciences, California Institute of Technology, 1200 E. California Blvd., Pasadena, CA 91125, USA}

\author[0000-0002-6618-1137]{W. Jerry Xuan}
\affil{Department of Astronomy, California Institute of Technology, 1200 E. California Blvd., Pasadena, CA 91125, USA}
\affil{Department of Physics and Astronomy, Pomona College, 333 N. College Way, Claremont, CA 91711, USA}

\begin{abstract}
Reference star differential imaging (RDI) is a powerful strategy for high contrast imaging. Using example observations taken with the vortex coronagraph mode of Keck/NIRC2 in $L^\prime$ band, we demonstrate that RDI provides improved sensitivity to point sources at small angular separations compared to angular differential imaging (ADI). Applying RDI to images of the low-mass stellar companions HIP~79124~C (192~mas separation, $\Delta L^\prime$=4.01) and HIP~78233~B (141~mas separation, $\Delta L^\prime$=4.78), the latter a first imaging detection, increases the significance of their detections by up to a factor of 5 with respect to ADI. We compare methods for reference frames selection and find that pre-selection of frames improves detection significance of point sources by up to a factor of 3. In addition, we use observations of the circumstellar disks around MWC~758 and 2MASS~J16042165-2130284 to show that RDI allows for accurate mapping of scattered light distributions without self-subtraction artifacts. 
\end{abstract}


\section{Introduction} \label{sec:intro}

Adaptive optics (AO) imaging surveys determine crucial occurrence statistics for giant exoplanet populations with masses $\gtrsim$1$~M_\mathrm{Jup}$ at orbital separations $>$10~au and provide unique insight into their formation and migration histories \citep{Bowler2016,Meshkat2017}. Furthermore, resolving low-mass companions from their host stars reduces measurement noise allowing for in-depth spectral characterization, which yields a wealth of information about the physical properties of their atmospheres \citep{Biller2017}. AO observations also allow for the detailed mapping of scattered light from dust in circumstellar disks providing context for planet formation theories.

Current and future infrared, high-contrast imaging programs seek to (1)~bridge the gap between the search completeness of direct imaging and radial velocity surveys, (2)~characterize the atmospheres of lower mass planets, and (3)~understand the interaction between planets and the circumstellar disks from which they form. All of the above science cases benefit from minimizing the angular separations from the host star at which instruments are sensitive enough to detect and characterize the planets and disks of interest. 

Reducing the inner working angle (i.e. the separation at which the transmission is 0.5) of ground-based, high-contrast imaging systems to their theoretical limits \citep[i.e. $\sim\lambda/D$, where $\lambda$ is the wavelength and $D$ is the telescope diameter;][]{Guyon2006} opens a critical search space at solar system scales (1-30~au) for stars within a few hundred parsecs. From a technical point of view, it will also pave the way to studies of temperate exoplanets in reflected light with future large aperture telescopes ($D\sim$30~m), especially in the habitable zones of nearby, late-type stars where the planet-to-star flux ratio is expected to be $\sim$10$^{-8}$ for planet radii of $\sim 1 R_\Earth$ \citep[see e.g.][]{Guyon2018}. Looking further into the future, these developments will also help maximize the number of Earth-like planets orbiting solar-type stars available for study with future exoplanet imaging space missions, such as the Habitable Exoplanet Observatory \citep[HabEx;][]{habexInterimReport} and Large UV/Optical/IR Surveyor \citep[LUVOIR;][]{LUVOIRinterimReport} concepts.

The vortex coronagraph mode \citep{Foo2005,Mawet2005} of the NIRC2 instrument at W.M. Keck Observatory \citep{Serabyn2017,Mawet2017,Ruane2017,Huby2017,Guidi2018,Reggiani2018,Xuan2018} has the optical throughput needed to search for giant planets down to $\lambda/D$, or $\sim$100~mas, for $L^\prime$ and $M_s$ bands ($\lambda$=3.4-4.8~$\mu$m). However, pairing the Keck adaptive optics system with a small inner working angle coronagraph does not immediately provide sensitivity to point sources with $\Delta$mag$\gtrsim$3 within 0\farcs5 of the host star in our experience. Surpassing this level also requires optimized observing strategies and post-processing \citep{Mawet2012}.

\begin{figure}
    \centering
    \includegraphics[width=\linewidth]{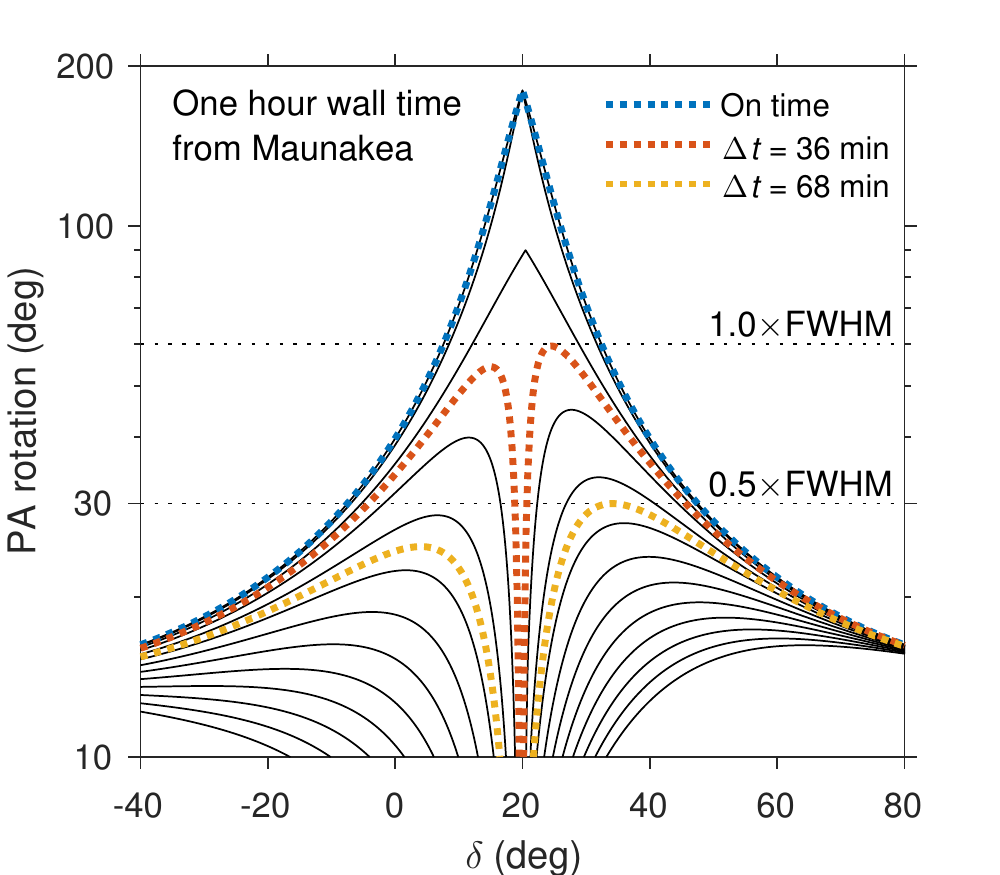}
    \caption{Parallactic angle rotation during a one hour observation from Maunakea, Hawaii (latitude 19.8$^\circ$ N) as function of target declination $\delta$. The dashed blue line represents an observation initiated exactly 30~min before the object crosses the meridian. The thin black lines represent observation windows shifted by $\Delta t$=15~min. The other thick dashed lines indicate the observation with maximum allowable timing error to achieve a PA rotation that moves a planet by an arclength equal to 1.0$\times$FWHM and 0.5$\times$FWHM at an angular separation of $\lambda/D$. Specifically, achieving 60$^\circ$ and 30$^\circ$ PA rotations at the most favorable $\delta$ requires the timing error to be $<$36~min and $<$68~min, respectively.}
    \label{fig:PArot}
\end{figure}

\subsection{The limitations of ADI}

High contrast imaging surveys with Keck/NIRC2 traditionally make use of angular differential imaging \citep[ADI,][]{Marois2006} in the search for giant planets, such as the four planets discovered orbiting HR 8799 \citep{Marois2008,Marois2010,Konopacky2016}. In this strategy, the telescope's field rotator is set to vertical angle mode, which keeps the beam fixed in azimuthal orientation with respect to the elevation axis and detector while the field of view rotates over the course of an observation. This assists in differentiating between true astrophysical objects, which revolve in the image, and starlight that leaks through the coronagraph, which typically leaves a residue of speckles in the image.

Although ADI is the most successful observational strategy for detecting giant planets at wide separations to date \citep[see recent review by][]{Chauvin2018}, it has some innate limitations. In order to clearly identify companions, each target must be observed for long enough to allow enough parallactic angle (PA) rotation such that a companion would move by a substantial fraction of the width of the point spread function (PSF). For instance, at the inner working angle of a vortex coronagraph (0\farcs08-0\farcs1 for NIRC2 $L^\prime$ and $M_s$ bands), the minimum desired PA rotation is $\sim$30$^\circ$, which moves the PSF of a companion by an arclength of roughly half of its full width at half maximum (FWHM). However, due to practical limitations, observations from ongoing surveys with Keck/NIRC2 vortex coronagraph have a median PA rotation of $\sim$11$^\circ$ \citep{Xuan2018}. 

To illustrate the problems associated with scheduling ADI observations, Figure~\ref{fig:PArot} shows the PA rotation achieved from Maunakea, Hawaii (latitude 19.8$^\circ$ N) in a one hour observation versus the target's declination and the timing of the observation. An ``on-time" observation (thick, blue, dashed line) in this case starts 30~min before the target crosses the meridian. The thin black lines shows how the PA rotation degrades when observation is initiated too early or too late by multiples of $\Delta t$=15~min. 

Figure \ref{fig:PArot} has several important implications: 

\begin{itemize}
    \item \textbf{Timing constraints}: ADI observations must be carried out during a relatively narrow time window to achieve a PA rotation needed to avoid self-subtraction effects~\citep{Marois2006}. For example, in order for a planet to move 30$^\circ$, or roughly an arclength of 0.5$\times$FWHM at an angular separation of $\lambda/D$, an observation at the most favorable $\delta$ must be initiated within $\pm$68~min of the optimal time. Increasing the path length requirement to 1.0$\times$FWHM (i.e. a PA rotation of 60$^\circ$) reduces the allowable timing error to $\pm$36~min. The timing requirement is even more strict at other $\delta$ values.
    \item \textbf{Limited sky coverage}: Although targets with declination $>$-40$^\circ$ are observable from Maunakea, achieving a PA rotation of $>$30$^\circ$ during a one hour observation is only possible on targets between $\delta$=-8$^\circ$ and $\delta$=48$^\circ$ (assuming the observations are timed perfectly). For PA rotation $>$60$^\circ$, the $\delta$ range becomes 8$^\circ$ to 32$^\circ$, ruling out more than 90\% of the sky.
    \item \textbf{Inefficient surveys of star forming regions}: Many of the most attractive targets for high-contrast imaging in the infrared are in nearby star forming regions (distances 120-150~pc), such as the Taurus and $\rho$ Ophiuchi molecular clouds \citep[see e.g.][]{Bowler2016}. However, these regions only extend over a few square degrees, allowing for one or two ADI sequences on these targets per night. Therefore, an ADI survey of these regions would need to be carried out over many nights.
    \item \textbf{Limited ``effective" inner working angle}: Unless the star falls within the declination ranges described above and the observations are well-timed, self-subtraction effects at small angular separations~\citep{Marois2006} ultimately limit the sensitivity near to the star and inner working angle of the coronagraph. 
    \item \textbf{Erases rotationally symmetric circumstellar disks}: Imaging scattered light from close to pole-on protoplanetary, transitional, and debris disks with ADI may not be possible because rotationally symmetric features self-subtract in post-processing and complex dust disk distribution become confounded by artifacts \citep{Milli2012}.
\end{itemize}

\subsection{The limitations of SDI}

An alternate speckle subtraction method known as spectral differential imaging \citep[SDI;][]{SparksFord2002} is applied on instruments with integral field spectrographs, such as Palomar/P1640 \citep{Hinkley2011}, Gemini/GPI \citep{Macintosh2014}, VLT/SPHERE \citep{Vigan2016}, and Subaru/SCExAO/CHARIS \citep{Jovanovic2015,Groff2017}. SDI decouples speckles from astrophysical objects based on their wavelength dependence rather than azimuthal rotation and therefore alleviates some of the aforementioned limitations of ADI. Between the shortest and longest wavelengths, speckles move radially by $\alpha\times\Delta\lambda/\lambda$, where $\alpha$ is the angular separation in units of $\lambda/D$. For instance, a radial change in the position of $>\lambda/D$ is only seen at angular separations $>$5~$\lambda/D$ using a typical filter with $\Delta\lambda/\lambda\approx0.2$. Therefore, SDI also suffers from self-subtraction effects and is not ideal for imaging close-in point sources or pole-on circumstellar disks. Additionally, Keck/NIRC2 does not have an integral field spectrograph for this purpose, making contemporaneous SDI impossible.

\subsection{Reference star differential imaging}

Reference star differential imaging \citep[RDI;][]{Lafreniere2009,Soummer2011,Gerard2016} is an alternative approach that uses images of other stars to build a model of the stellar PSF. RDI is a commonly used observational strategy for Hubble Space Telescope imaging observations of debris disks \citep{Golimowski2006,Schneider2009,Schneider2014,Choquet2016} and vortex coronagraph observations with the Hale telescope at Palomar Observatory \citep{Mawet2010,Serabyn2010,Mawet2011} where ADI is not possible due to its equatorial mount. In addition, RDI will very likely be applied for high-contrast imaging with future space telescopes, including the James Webb Space Telescope \citep[JWST;][]{Green2005} and the Wide Field Infrared Survey Telescope \citep[WFIRST;][]{Spergel2015,Bailey2018} as well as the HabEx \citep{habexInterimReport} and LUVOIR \citep{LUVOIRinterimReport} mission concepts. The latter two may push to very small angular separations using vortex or other small inner working angle coronagraphs \citep{Ruane2018_JATIS,Pueyo2017}. 

Here, we demonstrate how RDI with Keck/NIRC2 mitigates practical issues associated with ADI (and potentially SDI on other instruments). We demonstrate the benefits of the RDI observing strategy for ground-based imaging studies, especially the detection of point sources at small angular separations and mapping of scattered light from circumstellar disks. In the following sections, we detail the RDI method and strategy (Sec.~2) using three example observing nights that illustrate the benefits of RDI for point source detection (Sec.~3) and for imaging of circumstellar disks (Sec.~4) with Keck/NIRC2. Section~5 discusses our findings and the limitations of RDI, including error sources and potential artifacts. Section~6 summarizes our conclusions.

\section{RDI method and strategy}

RDI requires the observer to decide which reference frames and algorithms to use in order to model the stellar PSF. Reference frames may be specific reference stars purposely taken on the same night as the primary observation \citep[e.g.][]{Ruane2017}, frames from other stars that were coincidentally observed on the same night in the same observing mode \citep[e.g.][]{Xuan2018}, or from an archive of frames across observing programs \citep[e.g.][]{Choquet2016}. Example PSF reconstruction algorithms include subtracting a scaled version of the reference image \citep{Schneider2009}, principal components analysis \citep[PCA;][]{Soummer2012}, and non-negative matrix factorization \citep[NMF;][]{Ren2018}.

\cite{Xuan2018} showed that using all frames from a given night (excluding the target of interest) as reference frames and applying PCA improves the sensitivity to point sources at small angular separations with respect to ADI for many of the targets in our database of Keck/NIRC2 vortex coronagraph observations. In the following, we will discuss improvements to this strategy by combining this approach with frame pre-selection. 

For each observation, we apply basic pre-processing steps (see Appendix \ref{sec:preproc} for details) and PCA to estimate and subtract the starlight from the images using the Vortex Image Processing (VIP) software package \citep{GomezGonzalez2017}. Following \citet{Soummer2012}, a single frame may be written as $X = I  + A$, where $I$ is the stellar PSF after the coronagraph (a speckle field) and $A$ is the planet PSF. The stellar PSF is
\begin{equation}
    I = \sum_{k=1}^K \left< X, Z^{(k)} \right> Z^{(k)},
\end{equation}
where $\{Z^{(k)}\}_{k=1,...,N}$ are a basis set of images derived from a series of $N$ reference frames $\{R^{(k)}\}_{k=1,...,N}$, $\left< X, Z^{(k)} \right>$ is the projection of the frame onto the $k$th basis image, and $K$ represents the number of basis vectors used for the PSF model with $K\le N$. Here, the basis modes are computed using PCA.
In the case of point sources, we empirically determine the number of basis vectors, $K$, that provides the best sensitivity to fake companions injected in post-processing. Whereas in ADI-processed images the value of $K$ can drastically affect the sensitivity to point sources~\citep{Pueyo2016} and the occurrence of disk artifacts~\citep{Milli2012}, RDI results tend to be relatively insensitive to the choice of $K$ \citep{Soummer2012}. Rather, the choice of which reference frames to use to model the stellar PSF has a dominant effect on the quality of the final image.

\begin{deluxetable*}{llcccccccc}[t]
\tablecaption{Keck/NIRC2 vortex observations in $L^\prime$ band on UT 2016 Apr 13 \label{tab:HIPobs}}
\tablecolumns{10}
\tablenum{1}
\tablewidth{0pt}
\tablehead{
\colhead{Star name} & \colhead{Alt. Name} & \colhead{RA} & \colhead{Dec} & \colhead{Spec. Type} & \colhead{$N_\mathrm{frames}$} & \colhead{$t_\mathrm{int}$ (min)} &
\colhead{PA rot.} & \colhead{$V$} & \colhead{$W1^a$}
}
\startdata
HIP 41152 & HD 70313 & 08 23 48.5 & +53 13 11.0 & A3V & 30 & 10.0 & 12.9$^\circ$ & 5.54 & 5.21 \\
DX Leo & HD 82443 & 09 32 43.8 & +26 59 18.7 & G9V & 30 & 15.0 & 55.0$^\circ$ & 7.01 & 5.15 \\
TW Hya & TWA 1 & 11 01 51.9 & -34 42 17.0 & K6V & 70 & 35.0 & 21.6$^\circ$ & 10.5 & 7.10 \\
V1249 Cen & TWA 25 & 12 15 30.7 & -39 48 42.6 & M0.5 & 30 & 15.0 & 7.91$^\circ$ & 11.2 & 7.26 \\
V1252 Cen & TWA 10 & 12 35 04.2 & -41 36 38.5 & M2V & 30 & 15.0 & 7.82$^\circ$ & 13.0 & 8.09 \\
HIP 66704 & HD 119124 & 13 40 23.2 & +50 31 09.9 & F7.7V & 30 & 15.0 & 17.9$^\circ$ & 6.32 & 4.92 \\
HIP 78233 & HD 142989 & 15 58 29.3 & -21 24 04.0 & F0V & 36 & 18.0 & 32.4$^\circ$ & 9.10 & 7.63 \\
HIP 79124 & HD 144925 & 16 09 02.6 & -18 59 44.0 & A0V & 23 & 11.5 & 16.9$^\circ$ & 7.78 & 6.96 \\
\multicolumn2l{2MASS J16430128-1754274} & 16 43 01.3 & -17 54 27.4 & M0.5 & 30 & 15.0 & 11.8$^\circ$ & 12.6 & 8.44 \\
\multicolumn2l{2MASS J17520294+5636278} & 17 52 02.9 & +56 36 27.8 & M3.5V & 30 & 15.0 & 11.4$^\circ$ & 13.3 & 8.20 \\
V4046 Sgr & HD 319139 & 18 14 10.5 & -32 47 34.4 & K5+K7 & 20 & 10.0 & 5.63$^\circ$ & 10.7 & 7.12 \\
\enddata
\tablecomments{$^a$Wise~$W1$ mag \citep[3.4~$\mu$m;][]{Wright2010}. }
\end{deluxetable*}

\begin{figure*}[t!]
    \centering
    \includegraphics[width=0.9\linewidth]{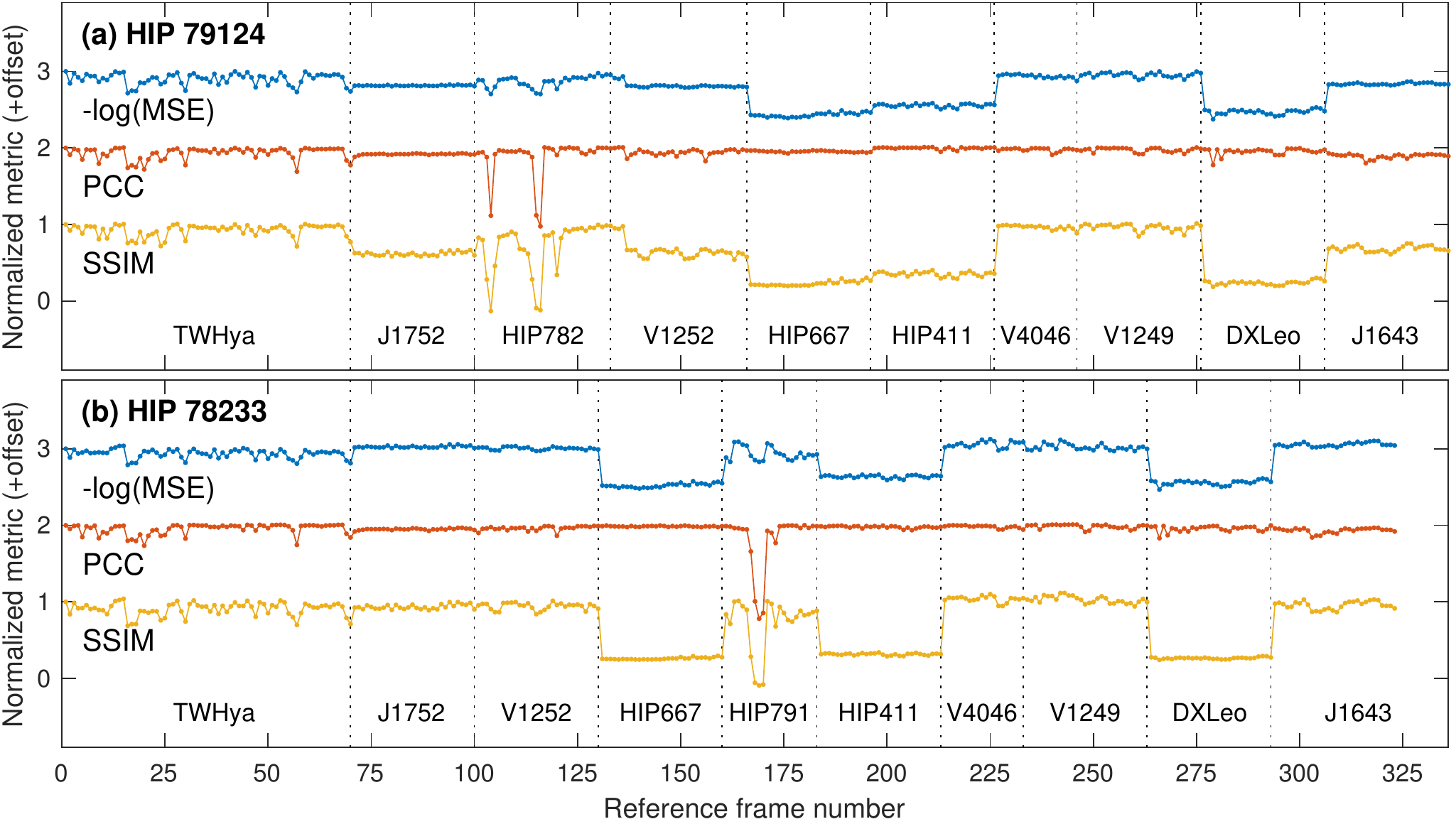}
    \caption{The MSE (shown as negative logarithm), PCC, and SSIM for all frames from UT~2016~Apr~13 (grouped by target in an arbitrary order) compared to the median science frame of the science observations: (a)~HIP~79124 and (b)~HIP~78233. MSE and PCC responds to differences in the image counts and structure, respectively; SSIM is correlated with both.}
    \label{fig:refFrameMetrics}
\end{figure*}

\begin{deluxetable*}{lccccccc}
\tablecaption{Best fit photometry and astrometry for the detected companions. \label{tab:HIPfits}}
\tablecolumns{8}
\tablenum{2}
\tablewidth{0pt}
\tablehead{
\colhead{Companion} & \colhead{Dist. (pc)$^a$} & \colhead{$\Delta L^\prime$} & \colhead{Sep (mas)} & \colhead{PA ($^\circ$)} & \colhead{$F/\sigma_F$} & \colhead{Mass ($M_\mathrm{Jup}$)$^b$} & \colhead{Proj. sep (au)} 
}
\startdata
HIP 79124 B & 132$\pm$1 & 2.49$\pm$0.04 & 971.5$\pm$1 & 100.61$\pm$0.03 & $>$31 & 630$\pm$17 & 128$\pm$1 \\
HIP 79124 C & 132$\pm$1 & 4.01$\pm$0.03 & 192$\pm$2 & 246.5$\pm$0.8 & 36 & 199$\pm$4 & 25.3$\pm$0.3 \\
HIP 78233 B & 256$\pm$4 & 4.78$\pm$0.12 & 141$\pm$3 & 6$\pm$1 & 8.6 & 198$\pm$19 & 36.0$\pm$0.9 \\
\enddata
\tablecomments{$^a$\citet{BailerJones2018}. $^b$Lower bound on mass; assumes the
BT-Settl model \citep{Allard2012} and an age of 10 Myr.}
\end{deluxetable*}

We explored a number of different methods for reference frame pre-selection. In each case, we compared the potential reference frames, $X_i$, with the temporal median of the science frames, $M$, over a speckle noise dominated region of the image (typically within 5-10~$\lambda/D$). We assigned a score to each reference frame using three metrics: the mean square error (MSE), the Pearson correlation coefficient (PCC), and the structural similarity index metric (SSIM) as defined in \citet{Wang2004}. The MSE of the $k$th reference frame is given by
\begin{equation}
    \mathrm{MSE}^{(k)} = \frac{1}{N_\mathrm{pix}}\sum_{i=1}^{N_\mathrm{pix}} \left(X_i^{(k)} - M_i\right)^2,
\end{equation}
where $N_\mathrm{pix}$ is the number of pixels in the comparison region. The PCC is 
\begin{equation}
    \mathrm{PCC}^{(k)}=\frac{\mathrm{cov}(X^{(k)},M)}{\mathrm{std}(X^{(k)})\mathrm{std}(M)},
\end{equation}
where $\mathrm{cov}(.)$ and $\mathrm{std}(.)$ represent the covariance and standard deviation. Specifically, the covariance is given by
\begin{equation}
    \mathrm{cov}(P,Q) = \frac{1}{N_\mathrm{pix}-1}\sum_{i=1}^{N_\mathrm{pix}}(P_i-\bar{P})(Q_i-\bar{Q}),
\end{equation}
where $P$ and $Q$ are the images being compared, with pixel-wise means $\bar{P}$ and $\bar{Q}$. The standard deviation is 
\begin{equation}
    \mathrm{std}(P) = \sqrt{\frac{1}{N_\mathrm{pix}-1}\sum_{i=1}^{N_\mathrm{pix}}(P_i-\bar{P})^2}.
\end{equation}
The final metric we consider is the mean SSIM:
\begin{equation}
    \mathrm{SSIM}^{(k)}= \frac{1}{N_\mathrm{pix}}\sum_{i=1}^{N_\mathrm{pix}} L_i^{(k)} C_i^{(k)} S_i^{(k)} ,
\end{equation}
where $L$, $C$, and $S$ are the luminance, contrast, and structural terms, which are computed over a FWHM$\times$FWHM window centered on pixel $i$. The luminance term is 
\begin{equation}
    L_i=\frac{2\;\bar{X} \bar{M} + c_1}{ \bar{X}^2 + \bar{M}^2 + c_1},
\end{equation}
the contrast term is 
\begin{equation}
    C_i=\frac{2\;\mathrm{std}(X) \mathrm{std}(M) + c_2}{ \mathrm{std}(X)^2 + \mathrm{std}(M)^2 + c_2},
\end{equation}
and the structural term is
\begin{equation}
    S_i=\frac{\mathrm{cov}(X,M) + c_3}{ \mathrm{std}(X)\mathrm{std}(M) + c_3}.
\end{equation}
$c_1$, $c_2$, and $c_3$ are small constants chosen to prevent instability when the denominator is otherwise close to zero. Qualitatively, $L_i$ is the relative change in luminance and responds similarly to MSE to differences in the pixel counts. $S_i$ is almost identical to the PCC. Thus, SSIM can be thought of as a mixture between the MSE and PCC metrics. 

In the following section, we investigate the performance gains achieved by modeling the stellar PSF using the principal components (PCs) of a set of reference frames selected based on their MSE, PCC, and SSIM values, as compared to the median science frame.


\section{Point source detection}
\label{sec:point_source}
To illustrate the utility of and trade-offs associated with using RDI for point source detection, we revisit Keck/NIRC2 observations taken during commissioning of the vortex coronagraph mode on UT~2016~Apr~13. Table \ref{tab:HIPobs} gives the full list of observations, consisting of 11 targets observed over the course of a full observing night for various scientific purposes. The targets span a large range of elevations, spectral types (Spec. type), and magnitudes, which we list in $V$ and Wise $W1$ bands \citep[3.4~$\mu$m;][]{Wright2010}. The observations also vary in the number of frames ($N_\mathrm{frames}$) and total integration times ($t_\mathrm{int}$). 

\citet{Hinkley2015} previously identified low-mass stellar companions orbiting at $\sim$20~au from two of the observed targets, HIP~79124 and HIP~78233, using aperture masking interferometry. HIP~79124 and HIP~78233 are classified as A0V and F0V stars, which reside in the Upper Scorpious subgroup of the Scorpius-Centaurus (USco) association at a distance of $\sim$120-150~pc \citep{deZeeuw1999} and, thus, have an estimated age of 5-20 Myr \citep{Pecaut2012,Song2012}. HIP~79124 is a triple system with an additional companion, HIP~79124~B, at $\sim$1$^{\prime\prime}$ from HIP~79124~A \citep{Lafreniere2014}. RDI allows us to directly image their close-in companions, HIP~79124~C and HIP~78233~B, even though their angular separation from the host star is $<$0\farcs2, or $2~\lambda/D$ in $L^\prime$ band. We use these previously reported astrometric measurements along with data presented in this study (Table~\ref{tab:HIPfits}) to check for common proper motion and confirm that HIP~78233~B, HIP~79124~B, and HIP~79124~C are all gravitationally bound to their host star.

Figure~\ref{fig:refFrameMetrics} shows the MSE, PCC, and SSIM for each possible reference frame from UT~2016~Apr~13, compared to the median frame in the HIP~79124 and HIP~78233 observations. As predicted, the MSE and PCC respond to different attributes, whereas the SSIM shows similar features to both MSE and PCC. In the case of HIP~79124 and HIP~78233, the brighter stars (HIP~66704, DX~Leo, HIP~41152) had considerably worse MSE and SSIM. Also, HIP~79124 and HIP~78233 were mutually poor reference stars for one another according to the PCC and the SSIM despite being observed consecutively. This is likely due to variable observing conditions at that point in the night.  

\begin{figure*}[t!]
    \centering
    \includegraphics[width=0.7\linewidth]{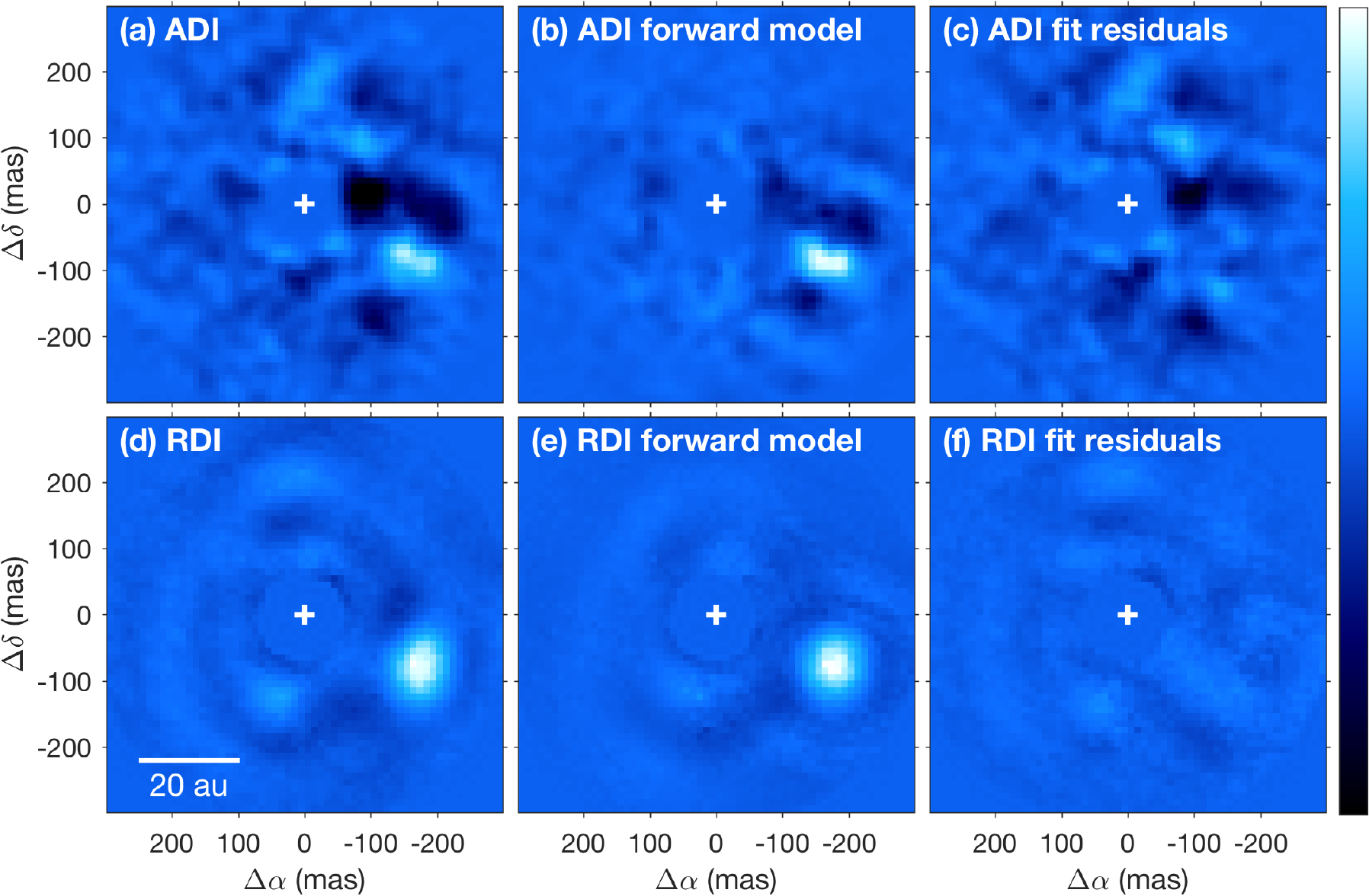}
    \caption{Observation of HIP 79124 C. (a) Image after subtracting stellar PSF using ADI. (b) Best forward model of the point source. (c) Residuals after subtracting the point source from the data. (d)-(f) Same as (a)-(c), but using RDI for the stellar PSF subtraction. North is up and East is left. The companion was retrieved at higher S/N in RDI compared to ADI. (a)-(c)~and (d)-(f)~are shown on the same scale. \label{fig:HIP79124}}
\end{figure*}
\begin{figure*}[t!]
    \centering
    \includegraphics[width=0.7\linewidth]{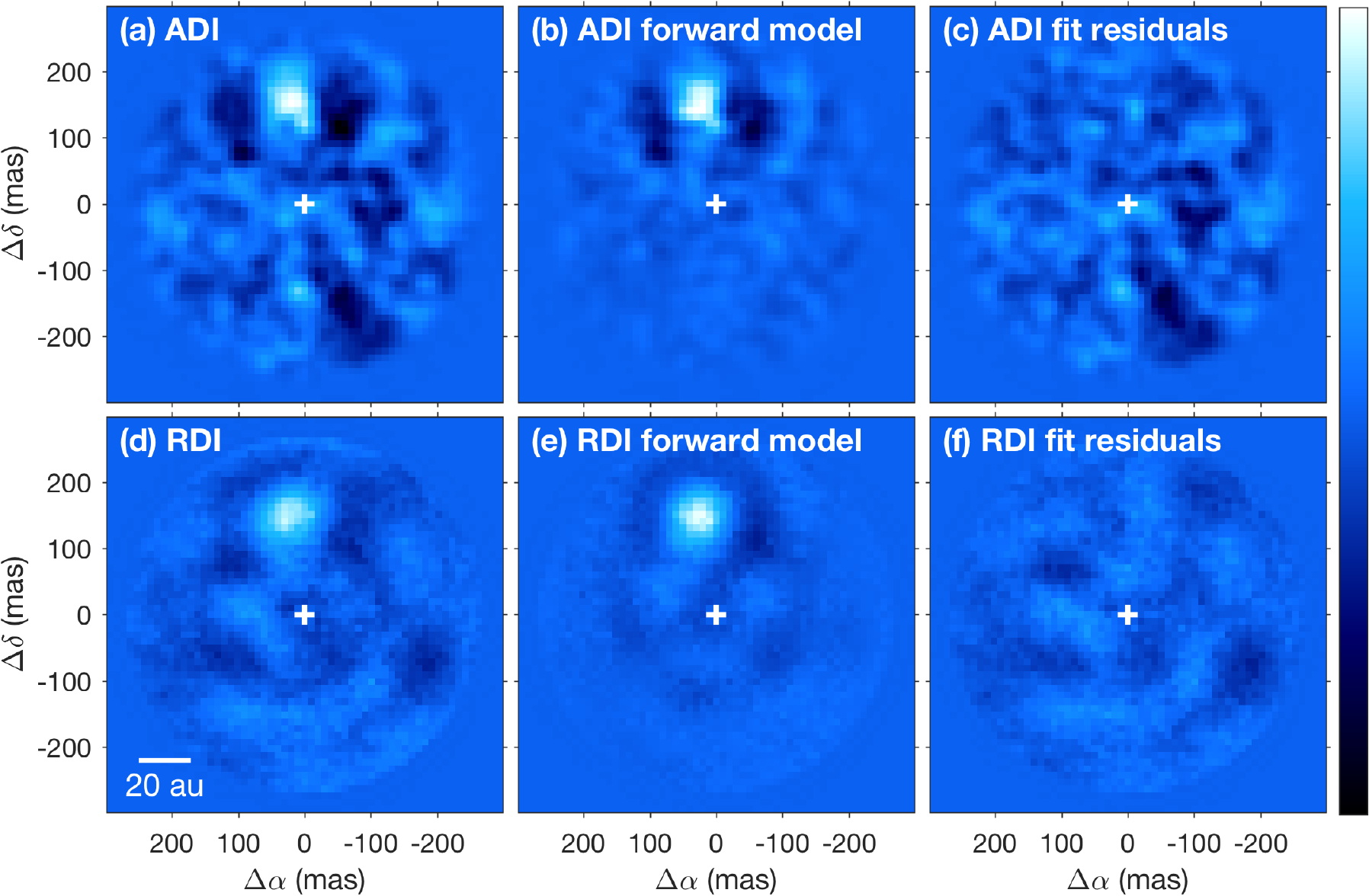}
    \caption{Same as Fig.~\ref{fig:HIP79124}, but for HIP 78233 B. \label{fig:HIP78233}}
\end{figure*}

Figures \ref{fig:HIP79124} and \ref{fig:HIP78233} show the ADI and RDI images of HIP~79124~C and HIP~78233~B, respectively. HIP~79124~C was previously imaged by \citet{Serabyn2017} using RDI. However, we have recovered HIP~78233~B for the first time. We show both examples here to demonstrate that, using PCA, the companions are detected at higher significance with RDI compared to our ADI reduction. 

\begin{figure*}[t!]
    \centering
    \includegraphics[width=0.8\linewidth]{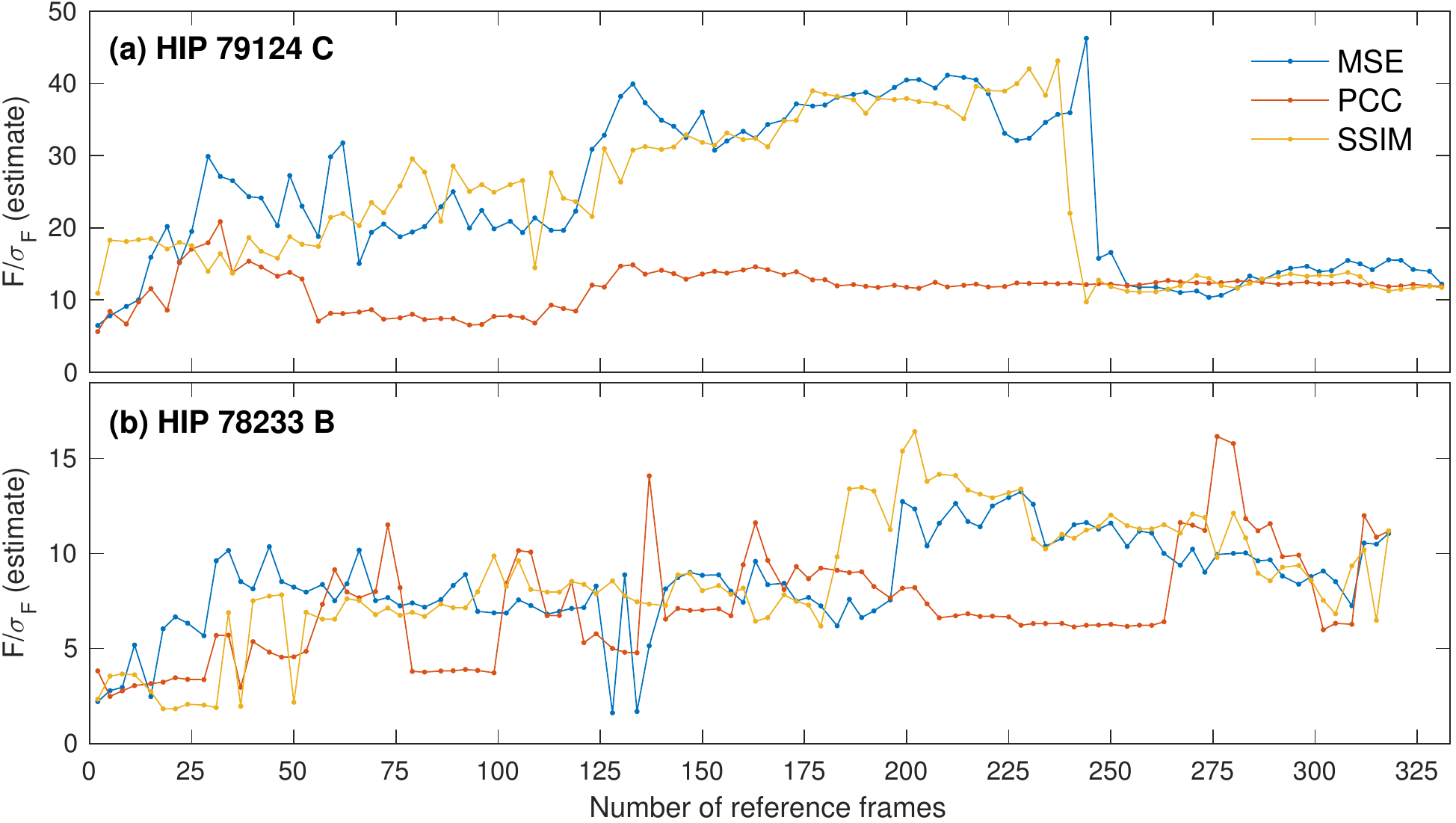}
    \caption{Estimated photometric $S/N=F/\sigma_F$ achieved with RDI versus the number of reference frames for the three frame selection methods. For the sake of finding the best set of reference frames, we injected a fake companion into the RDI residual cube with the best-fit flux and separation obtained using the maximum number of reference frames. }
    \label{fig:SNR}
\end{figure*}

In order to determine which reference frames to use, we ordered the reference frames according to the MSE, PCC, and SSIM and varied the number of frames from which we derive the PCs (see Fig.~\ref{fig:SNR}). In each case, we compute a forward model of the companion's PSF \citep[see e.g.][]{Lagrange2009,Soummer2012}. We subtracted a copy of the off-axis PSF at the location of the planet in each pre-processed science frame, varied its location and brightness, and repeated the PCA reduction until the values were minimized in the final images in a 2$\times$FWHM radius about the companion’s position using a downhill simplex algorithm. The forward model is defined as the difference between the original PCA reduction without injected or subtracted companions and the best-fit residuals.

The signal-to-noise ratio, $S/N$, is defined as the ratio between estimated flux, $F$, and the error in the flux fit, $\sigma_F$, as calculated using the method outlined in Appendix~\ref{sec:fwdmodeling}. We find that the MSE and SSIM metrics lead to higher $S/N$ for these objects compared to PCC, with SSIM offering a slight improvement. The optimal number of frames according to SSIM is 237 (out of 336) and 202 (out of 323) for HIP~79124~C and HIP~78233~B, respectively. We also optimized $K$ in this fashion for each number of reference frames considered, but found that $K$ has much less of an influence on the $S/N$. At the optimal number of reference frames, we used 11 and 15 PCs for HIP~79124~C and HIP~78233~B, respectively. Figures \ref{fig:HIP79124} and \ref{fig:HIP78233} show the result using the optimal combination of parameters. 
The $S/N$ has a steep drop when using more than $\sim$230 frames ordered by MSE and SSIM, which occurs when the reduction includes frames from the brightest stars (HIP~66704, DX~Leo, HIP~41152). This is expected since we found previously in \citet{Xuan2018} that the contrast in Keck/NIRC2 images is strongly correlated with stellar magnitude. 

Table \ref{tab:HIPfits} reports the astrometry and photometry values derived from the highest $S/N$ case. There are a few important differences between the values in Fig.~\ref{fig:SNR} and the final fits. First, we used the flux and position we retrieved using all reference frames to compute the fitting error in Fig.~\ref{fig:SNR}. In the final fits, we used the actual flux and position retrieved at the optimal number of reference frames and PCs for the injected, fake companion. Second, we applied a correction to the retrieved flux values to account for the throughput of the coronagraph, which was 81\% and 68\% for HIP~79124~C and HIP~78233~B, respectively, according to our optical model. In the end, we arrived at lower $S/N$ values for each companion than the estimates in Fig.~\ref{fig:SNR}. The final $S/N$ was better with RDI than ADI for both stars. Specifically, for HIP~79124~C, the $S/N$ in RDI was 36 whereas the best we could achieve in ADI was 7.9 with $K=2$ and a PA rotation of 17$^\circ$. Similarly, for HIP~78233~B, the $S/N$ in RDI was 8.6, but the best we achieved in ADI was 5.1 with $K=6$ despite having a PA rotation of 32$^\circ$.

Comparing to \citet{Hinkley2015}, the change in astrometry since 2010 is consistent with orbital motion, leading to the first confirmation that HIP~78233~B is bound. We updated the distances to HIP~79124 and HIP~78233 based on Gaia DR2 \citep{BailerJones2018}, which gave 132$\pm$1~pc and 256$\pm$4~pc for the respective stars, rather than the distances of 123~pc and 145~pc used in \citet{Hinkley2015}. Our measured flux ratio for HIP~79124~C, $\Delta L^\prime=4.01\pm0.03$, is inconsistent with both \citet{Hinkley2011} and \citet{Serabyn2017} by a small margin, who report $\Delta L^\prime=4.3\pm0.1$ and $\Delta L^\prime=4.2\pm0.1$, respectively. The increased flux estimate and updated stellar distance predicts a higher mass for both companions. Assuming the BT-Settl evolutionary model \citep{Allard2012} and an age of 10~Myr, we estimate minimum masses of 199$\pm$4~$M_\mathrm{Jup}$ and 198$\pm$19~$M_\mathrm{Jup}$, which are well above the hydrogen-burning limit.

For each point source companion, we use previously reported astrometry along with new data presented in this study to constrain the orbits of each companion around their host stars. Although two data points do not provide a well-constrained orbit, the long baseline provides some constraint on the actual semimajor axis and eccentricity. We use the ``Orbits of the Impatient'' (OFTI; \citealt{Blunt2017}) algorithm as implemented by the \verb+orbitize+ software package\footnote{\url{https://orbitize.readthedocs.io/}}. OFTI is a Bayesian rejection-sampling method that excels at finding constraints on long period orbits with sparse data points. We determine the median semimajor axis and 68\% confidence intervals to be $37^{+31}_{-12}$~au for HIP~78233~B, $157^{+82}_{-37}$~au for HIP~79124~B and $26^{+20}_{-8}$~au for HIP~79124~C. We also found that the data favors lower eccentricity orbits over high eccentricity orbits. Details of these orbit fits and the asymmetrical posteriors are discussed in Appendix \ref{sec:orbits}. The posteriors are also available online as supplementary data for use in other studies. 

\begin{deluxetable*}{llcccccccc}[t!]
\tablecaption{Keck/NIRC2 vortex observations in $L^\prime$ band on UT 2016 Oct 24.\label{tab:MWC758obs}}
\tablecolumns{10}
\tablenum{3}
\tablewidth{0pt}
\tablehead{
\colhead{Name} & \colhead{Alt. Name} & \colhead{RA} & \colhead{Dec} & \colhead{Type} & \colhead{$N_\mathrm{frames}$} & \colhead{$t_\mathrm{int}$ (min)} &
\colhead{PA rot.} & \colhead{$V$} & \colhead{$W1$} 
}
\startdata
TYC 1845-2048-1 & HD 284115 & 05 02 41.3 & +24 07 19.5 & K2 & 25 & 16.7 & 23.1$^\circ$ & 8.13 & 3.98$\pm$0.10 \\
MWC 758 & HD 36112 & 05 30 27.5 & +25 19 57.1 & A8V & 80 & 53.3 & 173$^\circ$ & 8.27 & 4.60$\pm$0.08 \\
TYC 1867-221-1 & HD 249769 & 05 58 47.2 & +25 15 28.8 & K7 & 50 & 33.3 & 171$^\circ$ & 8.83 & 4.14$\pm$0.09 \\
\enddata
\end{deluxetable*}

\begin{figure*}[t!]
    \centering
    \includegraphics[width=0.7\linewidth]{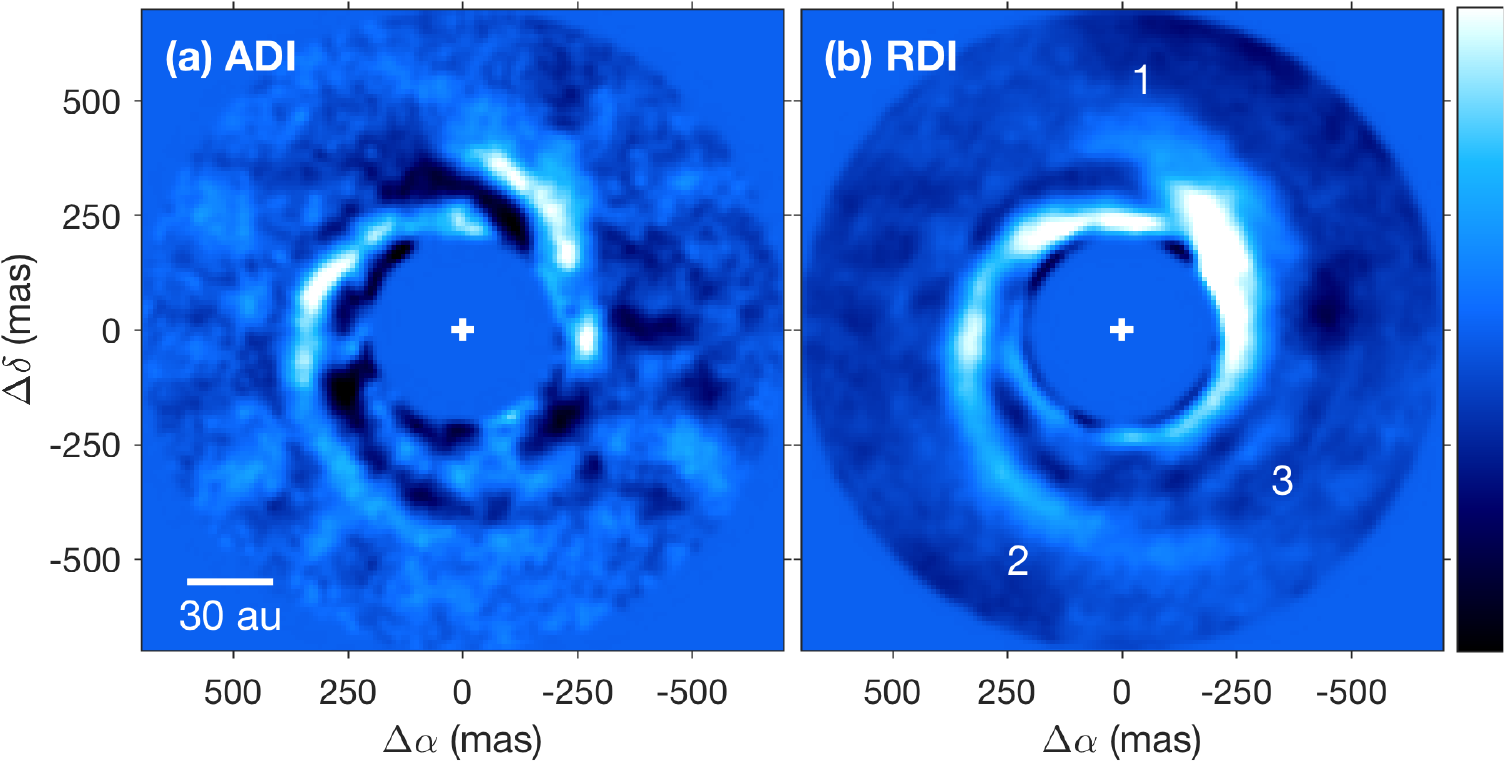}
    \caption{The circumstellar disk around MWC 758 after subtracting the stellar PSF using (a)~ADI and (b)~RDI. Labels `1,' `2,' and `3' indicate the prominent North and South spiral arms and the potential third arm, respectively.}
    \label{fig:MWC758}
\end{figure*}

In the two examples above, we have demonstrated that RDI offers significantly improved $S/N$ for companions at angular separations of $\lesssim 2~\lambda/D$ and that frame pre-selection is critical for maximizing the performance achieved with RDI. However, previous versions of the data reduction pipeline used by our team for Keck/NIRC2 vortex coronagraph observations \citep{Xuan2018} used all available reference frames from a given night by default. Although there is often an $S/N$ improvement at small separations using all frames over ADI, we are working towards using SSIM or similar metrics to perform the pre-selection of frames in a robust way and reprocessing archival NIRC2 data to reveal new point sources at close separations.

\section{Circumstellar disk imaging}

In addition to improving the sensitivity to point sources at small angular separations, RDI is also advantageous for disk imaging. In the following two examples, we use Keck/NIRC2 observations of circumstellar disks. In each case, reference stars were specifically chosen to be close in elevation when observed and to have similar brightness in $R$ and $L^\prime$ bands, where the wavefront sensor and science camera are sensitive.

\subsection{MWC~758}

MWC~758 (HD 36112) is a young \citep[3.5$\pm$2.0~Myr;][]{Meeus2012} stellar object in the Taurus star-forming region \citep[distance of 160~pc;][]{BailerJones2018} surrounded by a prominent protoplanetary disk with at least two massive spiral arms seen in scattered light observations~\citep{Grady2013,Benisty2015,Reggiani2018,Ren2018mwc758}. ADI images with Keck/NIRC2 from \citet{Reggiani2018} showed evidence of a potential third spiral arm. 

Here, we reprocessed one of the epochs presented in \citet{Reggiani2018} using reference stars observed directly before and after the previously published ADI sequence. Table \ref{tab:MWC758obs} gives the list of observations we used in the reduction. For ADI, we used all of the science frames and $K=6$. For RDI, we used similar methods as the previous section keeping the best 70\% (50 out of 75) of reference frames according to the SSIM metric and used the projection of 25 out of 50 PCs in the PSF model. However, since the reference stars were a good match to the science targets and conditions were stable during these observations, the variation in MSE, PCC, and SSIM was similar to that of one of the well-correlated observations in Fig.~\ref{fig:refFrameMetrics}. Therefore, the RDI image of the disk is visibly unchanged, barring some background variation, whether we use 10\% more or less frames or change the number of PCs by $\sim$20\%, while the ADI image is very sensitive to number of PCs. 

Figure \ref{fig:MWC758} shows the ADI and RDI images of MWC~758. In addition to the famous North and South spirals (labeled `1' and `2' in Fig.~\ref{fig:MWC758}b), we recovered the proposed third spiral arm feature (labeled `3') with both ADI and RDI for the first time. The ADI image is dominated by self-subtraction effects, while the RDI image is likely more representative of the true morphology of scattered light features. Although we have confirmed that the proposed third spiral arm was not an ADI artifact, it is unclear from the RDI image whether the feature is truly a third spiral arm or a continuation of the north arm wrapping almost 360$^\circ$ about the star. Nevertheless, this observation demonstrates that RDI is beneficial for mapping scattered light distributions in circumstellar disks without the sharpening of azimuthal features seen in the ADI images.

\subsection{J16042165-2130284}

We present an observation of the young stellar object 2MASS~J16042165-2130284 (from here on `J1604'). J1604 is a K2 star at a distance of 150~pc \citep{BailerJones2018} and a likely member of USco \citep{Carpenter2014} suggesting an age of 5-20 Myr \citep{Pecaut2012,Song2012}. J1604 is surrounded by a nearly pole-on \citep[7$^\circ$ inclination;][]{Mathews2012} pre-transitional disk~\citep{Espaillat2007}, the most massive known ($\sim0.1~M_\mathrm{Jup}$)  in USco \citep{Mathews2012}. Several previous works identified a ring-like disk in scattered light \citep{Mayama2012,Pinilla2015,Canovas2017,Pinilla2018}. This disk has received considerable attention owing to its variability \citep{Pinilla2015,Pinilla2018} and evidence of planet-induced dust filtration~\citep{Rice2006,Canovas2017}.


The observing list from UT~2017~May~10 (see Table~\ref{tab:J1604obs}) consists of seven observations, including J1604 and a reference star, 2MASS~J19121875-2137074 (from here on `J1912'), specifically selected prior to the observing night because its similar brightness and elevation. The remaining five targets were observed as part of an ongoing survey of M~stars (PI:~Mawet). The first three M~star targets were significantly worse reference stars according to their SSIM values because they were much fainter ($W1>9$) than J1604 ($W1=7.61$). The other two M~stars had SSIM values close to that of J1912.

\begin{figure*}[t!]
    \centering
    \includegraphics[width=0.7\linewidth]{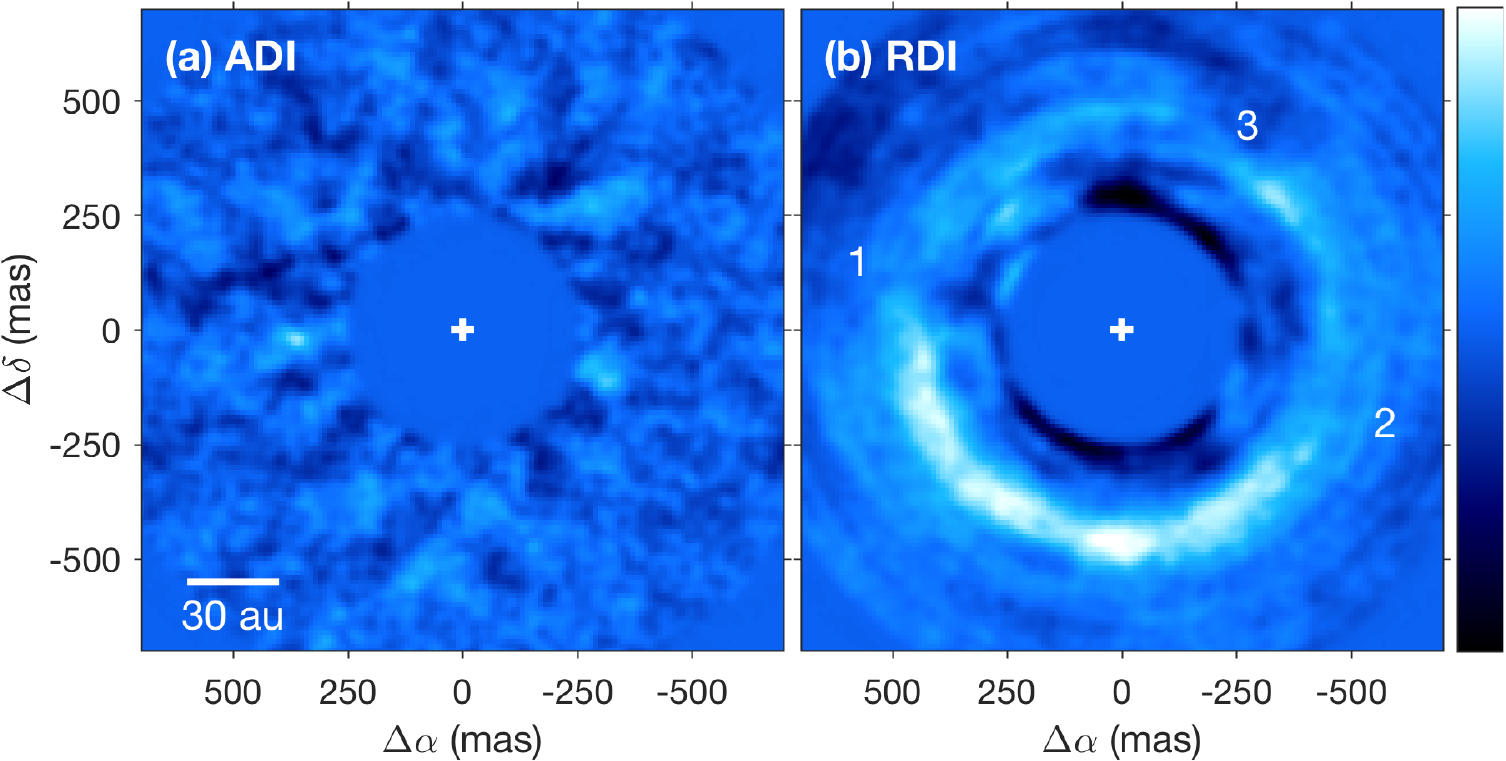}
    \caption{The 2MASS J16042165-2130284 disk after subtracting the stellar PSF using (a)~ADI and (b)~RDI. Labels `1,' `2,' and `3' indicate the the position of potential shadow features in the disk.}
    \label{fig:J1604}
\end{figure*}

Figure \ref{fig:J1604} shows the ADI and RDI images of J1604. Since the disk is nearly rotationally symmetric, the ADI image is consistent with noise for all values of $K$. However, the ring of scattered light is clearly visible in the RDI reduction. For the RDI image in Fig.~\ref{fig:J1604}b, we used the best 88 reference frames, according to their SSIM values, out of the 150 available and projected 44 PCs to build the stellar PSF model. Again, the morphology of the disk did not change much when changing the number of reference frames or PCs by $\sim20\%$ in this case. After removing poorly matching reference frames according to the SSIM, the projection coefficients for the first few PCs dominate.  

Polarized intensity maps of J1604 from \citet{Pinilla2018} show azimuthal dips attributed to shadowing by an inner disk. We see similar features labeled `1', `2', and `3' in Fig.~\ref{fig:J1604}b. These dips change with time indicating that the inner regions are highly dynamic, which may be evidence of a close-in massive companion. In the southern part of the disk, we estimate the inner and outer edges of the bright scattering feature appear at 440 and 500~mas, respectively. Compared to $J$-band (1.2~$\mu$m) observations taken with VLT/SPHERE by \citet{Pinilla2018} who measured a scattered light peak at 430~mas, the scattering surface appears slightly further from the star in $L^\prime$ band (3.8~$\mu$m), which may be evidence of spatial segregation of dust particle sizes and/or lower opacities at longer wavelengths. In future work, we will fit a forward model of the disk using radiative transfer and further investigate the wavelength dependence of the scattered light surface and shadowing effects in the disk (Wallack et al., in prep.), all of which would not be possible using ADI.

\begin{deluxetable*}{lcccccccc}[t!]
\tablecaption{Keck/NIRC2 vortex observations in $L^\prime$ band on UT 2017 May 10.\label{tab:J1604obs}}
\tablecolumns{9}
\tablenum{4}
\tablewidth{0pt}
\tablehead{
\colhead{Name} & \colhead{RA} & \colhead{Dec} & \colhead{Type} & \colhead{$N_\mathrm{frames}$} & \colhead{$t_\mathrm{int}$ (min)} &
\colhead{PA rot.} & \colhead{$V$} & \colhead{$W1$} 
}
\startdata
2MASS J11110358-3134591 & 11 11 03.6 & -31 34 59.1 & M & 20 & 15.0 & 6.99$^\circ$ & 14.4 & 9.35$\pm$0.02 \\
2MASS J11431742+1123126 & 11 43 17.4 & +11 23 12.6 & M & 20 & 15.0 & 10.1$^\circ$ & 12.3 & 9.12$\pm$0.03 \\
2MASS J13412668-4341522 & 13 41 26.7 & -43 41 52.2 & M3.5 & 20 & 15.0 & 6.18$^\circ$ & 14 & 9.70$\pm$0.02 \\
2MASS J16042165-2130284 & 16 04 21.7 & -21 30 28.5 & K2 & 73 & 54.8 & 36.2$^\circ$ & 11.9 & 7.61$\pm$0.03 \\
2MASS J18580415-2953045 & 18 58 04.2 & -29 53 04.5 & M0V & 20 & 15.0 & 7.25$^\circ$ & 11.8 & 7.86$\pm$0.02 \\
2MASS J19121875-2137074 & 19 12 18.8 & -21 37 07.4 & M & 50 & 37.5 & 19.8$^\circ$ & 11.3 & 7.42$\pm$0.03 \\
2MASS J20013718-3313139 & 20 01 37.2 & -33 13 13.9 & M1 & 20 & 15.0 & 6.95$^\circ$ & 12.3 & 8.14$\pm$0.02 \\
\enddata
\end{deluxetable*}

\section{Discussion}

The above sections exemplify that (1)~combining the small inner working angle vortex coronagraph and RDI is a sensitive technique for discovering faint companions $\sim$100~mas from the star and imaging scattering light from circumstellar disks with Keck/NIRC2 and (2)~optimizing the reference frame selection process leads to significant improvement in the quality of the stellar PSF subtraction. In this section, we describe the noise characteristics, artifacts, and limitations unique to this observing mode.

\subsection{Treatment of false positives}

Given the typical number of reference frames available per night and the limited diversity they provide, subtraction of the stellar speckle field is imperfect, and the residual speckles in RDI images tend to obey a highly non-Gaussian noise distribution \citep{Goodman1975,Aime2004}, especially within a few $\lambda/D$ of the star. However, the standard detection methods used in high-contrast imaging, including this work, estimate the standard deviation of the speckle noise, $\sigma_n$, as a function of angular separation from the star and set a fixed detection threshold (typically $5\sigma_n$), which leads to a substantial increase in the number of false positives compared to what is expected from normally distributed noise. In addition, since the spatial scale of speckles is roughly the FWHM of the off-axis PSF, $\sigma_n$ is estimated using only a few independent samples at small angular separations \citep{Mawet2014}. It is therefore especially challenging to differentiate between bright speckles and true companions in a single observation and to set robust upper limits on the brightness of unseen companions \citep[see e.g.][]{Ruffio2018} within a few $\lambda/D$ of the star. More work is needed to accurately model the residual speckle noise distribution given a set of science and reference frames. 

\subsection{Typical artifacts}

In addition to the non-Gaussian properties of the stellar speckle noise, there are other artifacts that may lead to false positives. For instance, several dust spots on the vortex focal plane mask in Keck/NIRC2 resemble point sources in RDI images because their brightness is not correlated with that of the star and background and therefore they are not well subtracted in post-processing. This requires the observer rule out detections corresponding to the location of the dust. We find that having a large amount of PA rotation helps blur these effects in the de-rotation step. 

The RDI approach assumes that stellar speckle noise dominates in both the science and reference frames. Generally speaking, RDI tends not to provide improved sensitivity at small angular separations on targets where spatially resolved circumstellar material dominates the speckle noise. In fact, projecting reference frames of point sources onto the science frames in an attempt to model the stellar PSF modifies the appearance of the circumstellar disk and generates additional speckle noise, which may obscure the true scattered light features and generate point-like false positives. Spatially resolved sources also have a mismatch in the radial size of the main stellar residuals that leaves behind a ring-like residual at the outer edge of the central lobe of the stellar PSF. A similar effect may occur for objects with vastly different colors, though we have not come across a clear example where chromatic effects dominate.

Lastly, there are speckles that appear in NIRC2 images at $\sim7~\lambda/D$ from the star whose brightness depends on the telescope elevation. We attribute these speckles to segment-to-segment phasing errors that depend on the direction of the gravity vector with respect to the primary mirror and the azimuthal position depends on the position of the rotator. For this reason, it may be beneficial to use reference stars at similar declination to the science target and similar instrument settings, including the rotator angle. 


\subsection{Sources of night-to-night variance}

The methods outlined above may be generalized to include reference frames from multiple observing nights. However, although the stellar PSF is relatively stable throughout an observing night, it tends to vary considerably from night-to-night for a number of reasons.  

The calibration of static wavefront error within NIRC2 uses a phase retrieval algorithm to reconstruct the wavefront from defocused images of the internal source. The solution changes on a nightly basis. Additionally, wavefront errors due to the primary mirror must be sensed and corrected by the Shack-Hartmann wavefront sensor in the Keck AO system \citep{Wizinowich2000}, but segment piston errors (or the ``terrace" modes) are not seen by the wavefront sensor and are therefore not corrected. This issue may be mitigated by the pyramid wavefront sensor under development at Keck Observatory \citep{Bond2018}.

For these reasons, reference frames are generally not well correlated from night-to-night. Nevertheless, there may be correlations with instrument settings (e.g. the rotator angle and wavefront sensor parameters) or observing conditions that can assist in frame pre-selection, which will be investigated in future work. 

\section{Conclusion}

Using the examples of HIP~79124~C and HIP~78233~B, we have demonstrated that RDI offers improved detection of point sources at small angular separation, by up to a factor of 5 with respect to ADI, further unlocking the potential of the vortex coronagraph on Keck/NIRC2. Furthermore, the observations of MWC~758 and 2MASS J16042165-2130284 show that RDI is beneficial for imaging of circumstellar disks because it preserves the morphology, which allows observers to directly interpret the imaged scattered light distribution. We find that frame pre-selection using image comparison metrics, such as the SSIM, significantly improves the performance of RDI, particularly for point source detection, astrometry, and photometry. RDI may help other current and future ground-based instruments achieve better performance at small angular separations, especially when used in conjunction with small inner working angle coronagraphs, and perform more efficient surveys by relaxing PA rotation requirements. RDI may also be the primary strategy for high-contrast imaging with future space missions, including JWST, WFIRST, HabEx, and LUVOIR. 

\acknowledgments

G. Ruane is supported by an NSF Astronomy and Astrophysics Postdoctoral Fellowship under award AST-1602444. 
The data presented herein were obtained at the W. M. Keck Observatory, which is operated as a scientific partnership among the California Institute of Technology, the University of California, and the National Aeronautics and Space Administration (NASA). The Observatory was made possible by the generous financial support of the W. M. Keck Foundation. The authors wish to recognize and acknowledge the very significant cultural role and reverence that the summit of Maunakea has always had within the indigenous Hawaiian community. We are most fortunate to have the opportunity to conduct observations from this mountain.

\facilities{Keck:II (NIRC2)}
\software{\texttt{VIP}~\citep{GomezGonzalez2017},\texttt{QACITS}~\citep{Huby2015,Huby2017}, \texttt{Astropy}~\citep{astropy}, \texttt{orbitize}~\citep{orbitize}}

\clearpage

\appendix

\section{Observing and image processing details}\label{sec:preproc}

In this section, we outline the details of our observations and the image processing steps used in our team's pipeline~\citep{Xuan2018}. All observations were carried out using the Keck~II telescope, natural guide star adaptive optics, the NIRC2 instrument, and the vortex coronagraph mode in $L^\prime$ band (3.7~$\mu m$). The angular resolution was 0.08$^{\prime\prime}$ and the plate scale was 0.01$^{\prime\prime}$ per pixel \citep{Service2016}. The field rotator was set to vertical angle mode, such that the telescope pupil tracks the elevation axis, to enable ADI. In addition to the science frames (see Tables \ref{tab:HIPobs}, \ref{tab:MWC758obs}, and \ref{tab:J1604obs}), we took several images of the off-axis PSF with a discrete integration time (DIT) of 0.008 seconds and 100 coadds as well as images of the sky background with integration times matching that of the science and off-axis PSF frames. The sky, off-axis PSF, and background frames were taken approximately every 30~min during the observing sequence. The alignment of the star and the center of the vortex focal plane mask was maintained by the QACITS tip-tilt control algorithm \citep{Huby2015,Huby2017}.


In the pre-processing stage, bad pixels identified in the dark frames and sky flats were replaced by the median of neighboring values in all frames. The sky flat was the median of 10 images of a blank patch of sky with the coronagraph focal plane mask removed (DIT of 0.75~s, 10 coadds each). We then subtracted sky background frames, taken with the coronagraph in place, from each frame individually using a scale factor to account for background variability. The frames were centered based on the position of the optical vortex core in the median of the science frames and each of individual science and reference frames were registered using the peak of the cross correlation with the median science frame. Our pipeline crops the raw 1024$\times$1024 pixel frames to 587$\times$587, which is the largest allowable square frame centered on the vortex focal plane mask. 

We applied PCA \citep{Soummer2012} to estimate and subtract the stellar contribution from each image using the VIP software package \citep{GomezGonzalez2017}. In the case of ADI, the PCs of a subset of pixels in the science frames make up the stellar PSF model. Although our standard pipeline uses a set of default annular reduction regions, we generally modify these regions to bound the companion or disk of interest on a case-by-case basis. For RDI, we derive the PCs and the stellar PSF model only from the reference frames. Each frame is de-rotated such that North is in the vertical direction and the ADI/RDI image is taken to be the temporal median of the de-rotated frames. We repeat these steps for all possible number of PCs. Later, an optimal number of PCs may be defined by maximizing the detection significance of a real or injected companion, depending on the scientific goal. For instance, an observer may wish to use a different number of PCs for point source detection and disk imaging.

\section{Error bars for astrometry and photometry of companions}\label{sec:fwdmodeling}


To estimate the errors in the position and flux estimates, with the best-fit companion subtracted from the data, we re-injected the PSF into the pre-processed science frames at the same separation and brightness, but varied the azimuthal angle and retrieved the photometry and astrometry using the same method for each injected companion tracing a full circle about the star. The step size in azimuthal angle was $360^\circ /(2\pi r)$, where $r$ is the radial position in units of the FWHM, to ensure that each fit was performed at the location of an quasi-independent speckle sample. We took the standard deviation of the measured flux and position of the injected companions as the error on each parameter. The signal-to-noise ratio, $S/N$, is defined as the ratio between estimated flux, $F$, and the uncertainty in the flux, $\sigma_F$. 

\section{Orbit fitting of point source companions with orbitize and OFTI}\label{sec:orbits}
In this section, we provide more details on the orbit fit and posteriors on the point source companions from the systems discussed in Section \ref{sec:point_source}. Table~\ref{tab:orbitize} summarizes the astrometry used in our fit. We use stellar distances from \citet{BailerJones2018} of 132$\pm$1~pc and 256$\pm$4~pc for HIP 78233 and HIP 79124, respectively. We use total system mass estimates from \citet{Hinkley2015}, corresponding to $1.7\pm0.1$~$M_\mathrm{Sun}$ for HIP 78233 and $2.48\pm0.45$~$M_\mathrm{Sun}$ for HIP 79124. We choose uninformed priors for all parameters. Namely, for the semimajor axis, we choose a prior probability distribution that is uniformly linear in log space (Jeffreys Prior); for inclination, we choose a prior probability distribution corresponding to $p(x) \propto sin(x)$; and for all other parameters, we choose a linearly uniform prior probability distribution. We fit the companions as point masses and for the HIP 79124 triple system, we treat each companion independently as their large mutual separation would make interactions between the B and C companions very small. The inclination convention is that $0^\circ \le i < 90^\circ$ orbits are counter-clockwise and $90^\circ \le i < 180^\circ$ are clockwise.

\begin{deluxetable}{cccc}
\tablecaption{Input astrometry for orbit fitting via \texttt{orbitize}\label{tab:orbitize}}
\tablecolumns{4}
\tablenum{5}
\tablewidth{0pt}
\tablehead{
\colhead{UT date} & \colhead{Separation (mas)} & \colhead{Position Angle ($^\circ$)} & \colhead{Reference}
}
\startdata
\sidehead{HIP 78233 B astrometry}
2010 Apr 26 & $133\pm3$ & $20\pm1$ & \citet{Hinkley2015}\\
2016 Apr 13 & $141\pm64$ & $6\pm1$ & this work\\
\sidehead{HIP 79124 B astrometry}
2008 May 25 & $990\pm1$ & $98.11\pm0.05$ & \citet{Lafreniere2014}\\
2016 Apr 13 & $971.5\pm1$ & $100.61\pm0.03$ & this work\\
\sidehead{HIP 79124 C astrometry}
2010 Apr 5 & $177\pm3$ & $242\pm1$ & \citet{Hinkley2015}\\
2016 Apr 13 & $192\pm2$ & $246.5\pm0.8$ & this work\\
\enddata
\end{deluxetable}

\begin{figure*}[t!]
    \centering
    \includegraphics[width=0.9\linewidth]{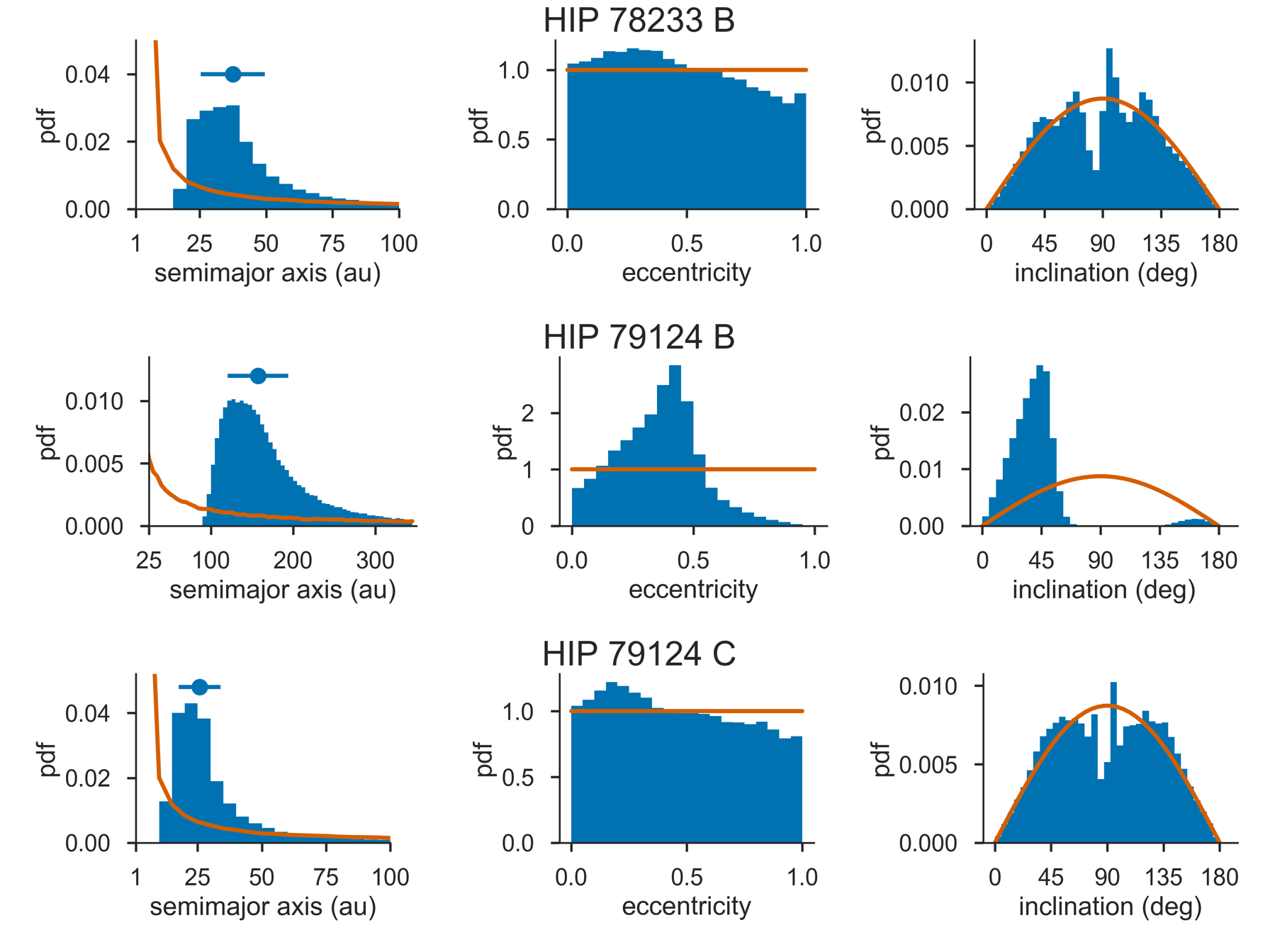}
    \caption{Marginalized probability density functions (pdfs) for three orbital parameters from orbit fits of three point source companions. The blue histogram shows the posterior pdf while the orange curve shows the prior pdf. For the semimajor axis pdf, the median and 68\% confidence region centered on the median is labelled as a point with horizontal bars. The posterior samples used to generate these pdfs are available as part of the online data.  \label{fig:orbitize}}
\end{figure*}

Using \texttt{orbitize}'s OFTI algorithm, we sample 100,000 orbits for each of the three point source companions and compute posterior probabilities. Figure~\ref{fig:orbitize} shows the marginalized posterior probabilities for orbital semimajor axis, eccentricity, and inclination. These figures also include the prior probability to demonstrate how well the data constrains the orbits. The marginalized posterior on semimajor axis is the best constraint on the companion's orbit size when accounting for all available astrometry. For HIP 78233 B and HIP 79124 C, while the eccentricity and inclination posteriors are not well constrained and are not very different from the prior. The eccentricity posterior distribution does show that lower values are more likely than higher values and the inclination posterior distribution disfavors a completely edge-on system (inclination of 90$^\circ$). For HIP 79124 B, the majority of the eccentricity distribution is at values below 0.5 and the inclination distribution strongly favors an inclination less than 90$^\circ$, indicating that the object is orbiting counter-clockwise from our viewing angle. These posteriors are available as part of this article's online data for use in further calculations as a table of orbit samples.

\clearpage


\end{document}